\documentclass[usenatbib,useAMS]{mn2e}
\usepackage{txfonts}
\usepackage{mathptmx}
\usepackage[T1]{fontenc}
\usepackage{ae,aecompl}
\usepackage{natbib}
\usepackage{multicol}

\usepackage{graphicx}	
\usepackage{amssymb}	
\usepackage{aecompl}
\usepackage{color}

\newcommand{\aj}{AJ}
\newcommand{\apj}{ApJ}
\newcommand{\apjl}{ApJ}
\newcommand{\apjs}{ApJS}
\newcommand{\mnras}{MNRAS}
\newcommand{\aap}{A\&A}
\newcommand{\araa}{ARA\&A}

\newcommand{\jqsrt}{JQSRT}

\newcommand       \mum          {\rm \mu m}
\newcommand       \simali       {\sim\,}

\title[Mid-IR polarimetry of WL 16]{The Mid-Infrared Polarization of 
       the Herbig Ae Star WL 16: An Interstellar Origin?}

\author[H. Zhang et al.]{Han Zhang$^{1}$\thanks{E-mail: hanzh0420@ufl.edu},
Charles M. Telesco$^{1}$,
Eric Pantin$^{1,2}$,
Dan Li$^{1}$,
Christopher M. Wright$^{3}$,
\newauthor
Naib\'{i} Mari\~{n}as$^{1}$,
Peter Barnes$^{1,4}$,
Aigen Li$^{5}$ and Christopher Packham$^{6}$
\\
$^{1}$Department of Astronomy, University of Florida, Gainesville, Florida, 32611, USA\\
$^{2}$Service d'Astrophysique CEA Saclay, France\\
$^{3}$School of Physical, Environmental and Mathematical Sciences, UNSW Canberra, PO Box 7916, Canberra BC 2610, Australia\\
$^{4}$School of Science and Technology, University of New England, Armidale, NSW 2351, Australia\\
$^{5}$Department of Physics and Astronomy, University of Missouri, Columbia, MO 65211, USA\\
$^{6}$Physics \& Astronomy Department, University of Texas at San Antonio, San Antonio, Texas, 78249, USA
}

\date{Accepted XXX. Received YYY; in original form ZZZ}

\pubyear{2016}

\begin{document}
\label{firstpage}
\pagerange{\pageref{firstpage}--\pageref{lastpage}}
\maketitle

\begin{abstract}
We present high-resolution (0$\farcs$4) mid-infrared (mid-IR) polarimetric images and spectra of WL 16, a Herbig Ae star at a distance of 125 pc. WL 16 is surrounded by a protoplanetary disk of $\simali$900 AU in diameter, making it one of the most extended Herbig Ae/Be disks as seen in the mid-IR. The star is behind, or embedded in, the $\rho$ Ophiuchus molecular cloud, and obscured by 28 magnitudes of extinction at optical wavelengths by the foreground cloud. Mid-IR polarization of WL 16, mainly arises from aligned elongated dust grains present along the line of sight, suggesting a uniform morphology of polarization vectors with an orientation of 33\degr (East from North) and a polarization fraction of $\simali$2.0\%. This orientation is consistent with previous polarimetric surveys in the optical and near-IR bands to probe large-scale magnetic fields in the Ophiuchus star formation region, indicating that the observed mid-IR polarization toward WL 16 is produced by the dichroic absorption of magnetically aligned foreground dust grains by a uniform magnetic field. Using polarizations of WL 16 and Elias 29, a nearby polarization standard star, we constrain the polarization efficiency, \textit{$p_{10.3}/A_{10.3}$}, for the dust grains in the $\rho$ Ophiuchus molecular cloud to be $\simeq$ 1.0\% mag$^{-1}$. WL 16 has polycyclic aromatic hydrocarbon (PAH) emission features detected at 8.6, 11.2, 12.0, and 12.7 $\mum$ by our spectroscopic data, and we find an anti-correlation between the PAH surface brightness and the PAH ionization fraction between the NW and SW sides of the disk.

\end{abstract}

\begin{keywords}
techniques: polarimetric --
ISM: dust, extinction -- 
ISM: magnetic fields --  
stars: formation -- stars: individual (WL 16, Elias 29)
\end{keywords}



\section{INTRODUCTION}\label{sec:I}
Magnetic fields (B-fields) play an important role in almost all stages of star formation as discussed in the comprehensive star formation review by \citet{2007ARA&A..45..565M}. However, there are still many uncertainties about how B-fields regulate the protoplanetary disk formation and evolution. For example, magnetically driven core-collapse models \citep{1987ARA&A..25...23S,1993ApJ...417..220G} predict an hourglass-shaped B-field geometry at early stages in the disk evolution, a scenario challenged by recent observations \citep[e.g.][]{2013ApJ...770..151C, 2011ApJ...732...97D, 2013ApJ...768..159H}. The only way to address these issues is with high angular resolution observations of the B-field morphologies in a variety of young disks and their environments.

Polarimetry is a potentially incisive observational probe of B-field morphology \citep{2015MNRAS.453.2622B,2012ARA&A..50...29C, 2009ApJS..182..143M, 2014ApJS..213...13H}. Dust grains can polarize light by scattering, dichroic absorption, or dichroic emission, the latter two processes attributed to non-spherical dust grains with their long axes aligned perpendicular to the B-field lines, perhaps as a result of radiative torque \citep{2007JQSRT.106..225L, 2014MNRAS.438..680H}. Dichroic absorption by aligned non-spherical dust grains can partially polarize background starlight with the transmitted E-field direction parallel to the aligning B-field lines \citep{2014ApJ...793..126C}. Light scattering by spherical and non-spherical dust grains can produce high fractional polarization prominently in the optical and near-IR. At far-IR and sub-mm wavelengths, aligned non-spherical dust grains emit polarized light with the emitted E-field direction being perpendicular to the B-field lines. In the mid-infrared (mid-IR, 8--30 $\mum$), the situation becomes more complicated, since the observed polarization can be a combination of dichroic absorption, emission, and/or scattering \citep{2005ASPC..343..293A}. Nevertheless, mid-IR polarimetry has some distinct advantages compared to other wavelength regions: the predicted emissive polarization is much larger than that at longer wavelengths \citep{2007ApJ...669.1085C}, and we can potentially disentangle both the absorptive and emissive polarization components simultaneously along the line of sight and thereby infer the three-dimensional structure of the B-fields \citep{2015MNRAS.453.2622B, 2000MNRAS.312..327S}.

Herbig Ae/Be stars (1.5$<$M$_{\star}$/M$_{\odot}$$<$8) are the higher-mass counterparts to pre-main-sequence low-mass T Tauri stars, with which they share some photometric and spectroscopic characteristics \citep{1960ApJS....4..337H, 1992ApJ...397..613H}. WL 16 was discovered by  \cite{1983ApJ...274..698W} and identified as a late-stage Herbig Ae star embedded in the $\rho$ Ophiuchus molecular cloud L1688 (\cite{2003ApJ...584..832R}, herafter RB03). Key physical properties of WL 16 are given in Table \ref{tab:1}. No outflow and only weak 1.3 mm emission are observed for this object (RB03). Because of its high extinction ($A_V\approx$ 25--30\,mag), WL 16 is undetectable in the optical but displays extended, resolved emission in the mid-IR. The most plausible interpretation of the extended emission is that it is a disk with a diameter of nearly 900 AU. Kinematic modeling of near-IR CO vibrational overtone emission arising from the innermost region suggests that there is indeed a flat Keplerian gas disk \citep{1993ApJ...411L..37C, 1996ApJ...462..919N}. Nevertheless, whether the entire structure is an unusually large protoplanetary disk or a smaller disk associated with a remnant of the collapsing envelope is still unclear. The very extended emission is generally ascribed to a population of very small grains (VSGs) and polycyclic aromatic hydrocarbons (PAHs). The latter population is indicated by the IR spectra of WL 16 rich in PAH emission features \citep{1998ApJ...504L..43D, 2007A&A...476..279G}. Because of their small size (several dozens to hundreds of carbon atoms), PAHs are dynamically well coupled to the gas and less affected by dust settling. The PAHs observable in the mid-IR likely reside in the disk surface layers. PAHs can also be easily ionized by stellar UV radiation due to their low ionization potentials, and ionized PAHs can be the tracer of low density optically thin regions \citep{2014A&A...563A..78M}. 

In this paper, we present high-resolution ($\sim$0$\farcs$4) mid-IR polarimetric imaging and spectropolarimetry of WL 16.  In addition, we find it illuminating to consider complementary observations of Elias 29 (our calibration star), which is a low-mass luminous Class I protostar (36 L$_{\odot}$) and a neighbor of WL 16 in the $\rho$ Ophiuchus molecular cloud \citep{2002ApJ...570..708B}. The paper is organized as follows. In Section \ref{sec:secII} we describe our observations. In Section \ref{sec:secIII}, we discuss the origin of the observed polarization from polarimetric imaging and spectroscopy, with detailed consideration given to the contribution associated with the foreground extinction and B-field. In Section \ref{sec:secIV} we examine the spatial distribution of the ionized and neutral PAHs along the major axis of the disk and draw attention to morphological features that may provide clues to WL 16's dynamical evolution. Finally, in Section \ref{sec:secV}, we summarize this work.

\begin{table}
\caption{Basic Properties of WL 16}
\label{tab:1}
\begin{tabular}{lcc}
\hline    
Properties                      &               Values   & References \\
\hline                    
Distance             &               125 pc  &  RB03 \\
Stellar mass                   &               4 $M_{\odot}$ &  RB03 \\
Luminosity   &              250 $L_{\odot}$ &  RB03\\
Disk inclination$^{a}$                    &                62.2\degr $\pm$ 0.4\degr &  RB03 \\
Disk PA$^{b}$                        &   60\degr $\pm$ 2\degr &  RB03 \\
Accretion rate     &           10$^{-6.8}$   $M_{\odot}$ yr$^{-1}$  & \citet{2006AA...452..245N}  \\  
Disk mass                                &       <0.001 $M_{\odot}$ & \citet{2007ApJ...671.1800A} \\
Disk diameter                                 &   900 AU & RB03 \\

\hline
\multicolumn{3}{l}{$^{a}$ 0\degr for face-on}\cr
\multicolumn{3}{l}{$^{b}$ Position angle of the major axis of the disk}\cr
\end{tabular}  
\end{table}

\section{OBSERVATIONS}\label{sec:secII}
CanariCam is the mid-IR (8--25 $\mum$) multi-mode facility camera of the 10.4-meter Gran Telescopio Canarias (GTC) on La Palma, Spain \citep{2003SPIE.4841..913T}.  It has a field of view of 26$\arcsec \times19\arcsec$ with a pixel scale of 0$\farcs$079, which provides Nyquist sampling at 8.7 $\mum$. In the polarimetry mode, the actual field of view is reduced to 26$\arcsec \times 2\farcs6$ after a focal mask (to avoid overlapping between \textit{o} and \textit{e} beams) is used.
We obtained polarimetric images of WL 16 in three filters near 10$\mum$  on 6 August 2013 and a low-resolution (R$\approx$50) polarimetric spectrum of WL 16 from 7.5 to 13$\mum$ on 4 and 5 July 2015 (Table \ref{tab:2}), as part of the CanariCam Science Team guaranteed time program (PI: C. M. Telesco) at the GTC.
The imaging and spectroscopic observations of WL 16 were interlaced with observations of standard star HD 145897 \citep{1538-3881-117-4-1864} for flux and point-spread-function (PSF) calibration, and the standard star Elias 29, selected from \cite{2000MNRAS.312..327S} to calibrate the polarization position angle (PA). The achieved angular resolution (full-width at half maximum intensity) for the polarimetric images was 0$\farcs$4--0$\farcs$6 (Table \ref{tab:2}).
CanariCam was rotated so that the polarimetry field mask and the detector array's long axis were along the major axis of the disk at PA=60$\degr$. The standard chop-nod technique was applied with a 15$\arcsec$ chop throw in the North-South direction. For spectropolarimetry, we positioned the 1$\farcs$04$\times$2$\farcs$08 slit with the slit's longer axis oriented at 72$\degr$ from the North to cover the brightest part of the disk. The chop throw was set at  8$\arcsec$ in the North-South direction.
 
\begin{figure}
\begin{center}
\includegraphics[width=\columnwidth]{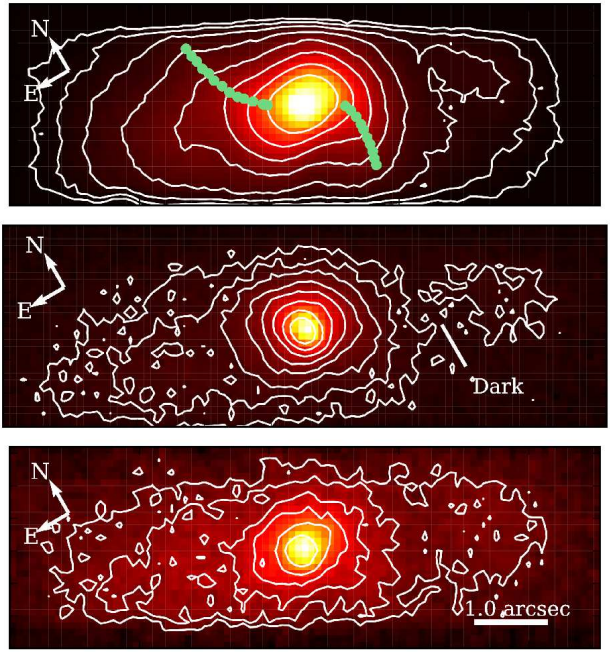} 
\caption{Total intensity maps of WL 16 at 8.7, 10.3, and 12.5 $\mum$ from top to bottom. 
Contours are surface brightness and logarithmically spaced. Parts of the disk structure are truncated by the CanariCam mask.  
Some peculiar features of the disk, including a spiral-arm-like structure (green dots) and a dark lane in the SW disk, which may indicate a disk warp, are highlighted.}
\label{fig:0}
\end{center}
\end{figure}

\begin{figure}
\begin{center}
\includegraphics[width=\columnwidth]{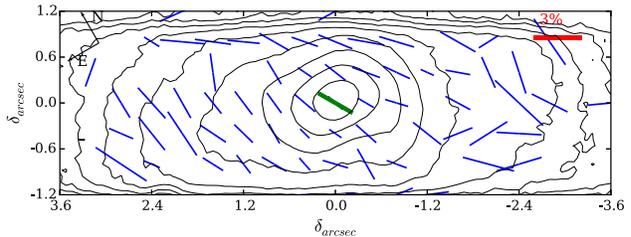} 
\caption{The 8.7-$\mum$ linear polarization map of WL 16 superimposed on (total intensity) contours.
The polarization vectors are plotted at the center of each 5 $\times$ 5 binned pixels, corresponding to 0$\farcs$4 $\times$ 0$\farcs$4 (50 $\times$ 50 AU), and are only plotted where $p/\sigma_{p}$ $\geqslant$ 2.0 and p $\leqslant$ 6\%. The lengths of polarization vectors are scaled to the polarization percentage and orientations correspond to the polarization PA. The green vector at the center, which is oriented at 30$\degr$ from North, is the polarization PA observed in the optical \citep{1990ApJ...359..363G} and near-IR \citep{1988MNRAS.230..321S} bands.}
\label{fig:1}
\end{center}
\end{figure}

The data were reduced using custom IDL software, as described in \cite{2014ascl.soft05014B, 2014ascl.soft11009L}. We computed normalized Stokes parameters \textit{q} and \textit{u}, where \textit{q=Q/I} and \textit{u=U/I}. The degree of polarization $p=\sqrt{q^{2}+u^{2}-\sigma_{p}^{2}}$, where the last term (the ``debias'' term) was introduced to remove the positive offset in the signal floor resulting from noise. The polarization PA is computed as $\theta=0.5{\rm arctan}(u/q)$. The uncertainties $\sigma_{q}$ and $\sigma_{u}$ associated with the normalized Stokes parameters were derived using a 3-sigma clipping algorithm \citep{1987igbo.conf.....R}, and they were then propagated through the analysis to obtain the polarization uncertainty $\sigma_{p}$ and polarization PA uncertainty $\sigma_{\theta}=\sigma_{p}/2$. The total intensity (Stokes \textit{I}) images of WL 16 in the three passbands are presented in Fig. \ref{fig:0}. The polarization image at 8.7 $\mum$, where the highest signal-to-noise ratio is achieved, is shown in Fig.\,\ref{fig:1}. 
Polarizations in the \textit{Si2} (8.7 $\mum$)  and \textit{Si4} (10.3$\mum$) filters measured with an aperture of 1$\arcsec$ in  radius  centered  on  the  star  are  given  in  Table \ref{tab:3}.  The \textit{Si6} (12.5 $\mum$) data are too noisy to provide meaningful polarization information.  

Note that in the polarimetric images, the upper and lower edges of the WL 16 disk are truncated by the focal mask, and the polarization vectors near the edges are unreliable and thus not displayed in Fig. \ref{fig:1}. The data presented in Fig.\,\ref{fig:1} have been smoothed by 5$\times$5 pixel (0$\farcs$4 $\times$ 0$\farcs$4) binning. Vectors are only plotted where $p/\sigma_{p}$ $\geqslant$ 2.0 and $p \leqslant$ 6\%.

\begin{figure}
\begin{center}
\includegraphics[width=\columnwidth]{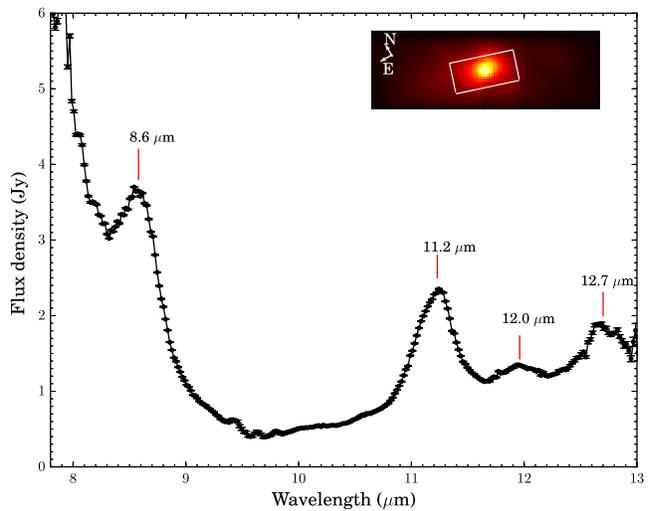} 
\caption{The low-resolution (R$\approx$50) spectrum of the brightest central 1$\farcs$6 (21 pixels) region of WL 16. The slit (white rectangle) is shown on the inset image of WL 16 at 8.7 $\mum$. 
The raw data were smoothed with a boxcar of 3 pixels (0.06 $\mum$) in width. 
PAH emission features are seen at 8.6, 11.2, 12.0, and 12.7 $\mum$.
The 8.6 $\mum$ feature originates from C-H in-plane bending. The 11.2, 12.0, and 12.7 $\mum$ features originate from C-H out-of-plane bending.}
\label{fig:2}
\end{center}
\end{figure}

\begin{figure}
\begin{center}
\includegraphics[width=\columnwidth]{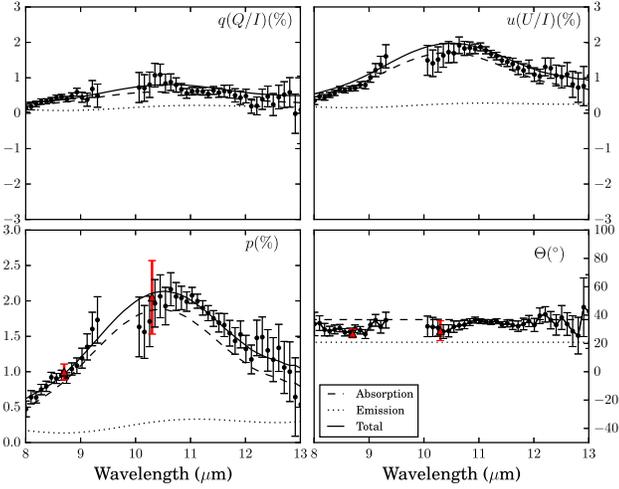}
\caption{
Polarization and emission/absorption decomposition results of WL 16. 
Black dots with 1 $\sigma$ error bars are measurements from spectropolarimetry. The atmospheric ozone band is masked out. The data are integrated across the central 1$\farcs$6 (21 pixels) region. 
We downsample the original data over 0.1 $\mum$ wavelength (5 pixels) bins and then apply a 5-pixel boxcar smoothing to increase the signal-to-noise ratio. The resultant spectral resolution is about 25.
Overlayed red points with 1 $\sigma$-error are measurements from polarimetric imaging at 8.7 and 10.3 $\mum$.
The dashed and the dotted lines are the best fitting absorptive and emissive components, and the solid lines are the combination.  
 From upper left to the bottom right: 1) Normalized Stokes parameter \textit{q (Q/I)}. (b) Stokes \textit{u (U/I)}. (c) Polarization degree \textit{p}. 
(d) Position Angle (PA) \textit{$\theta$}. $\theta$ is approximately 33$\degr$ regardless of the wavelength. The decomposition fitting indicates that it is absorptive-polarization dominant.}
\label{fig:3}
\end{center}
\end{figure}

\begin{figure}
\begin{center}
\includegraphics[width=\columnwidth]{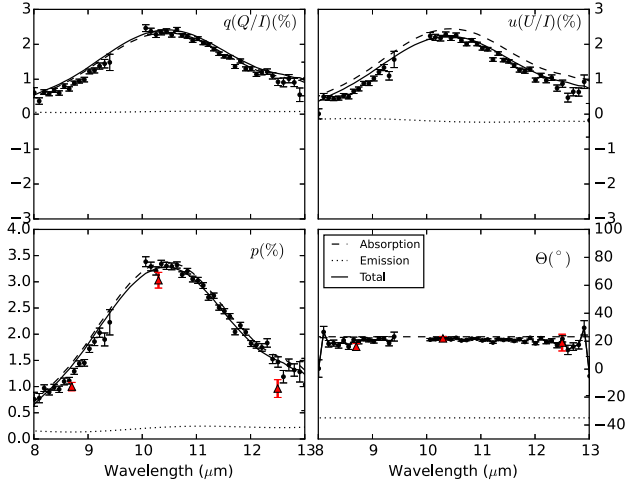}
\caption{Polarization and emission/absorption decomposition results of Elias 29. 
Black dots with 1 $\sigma$ error bars are measurements from spectropolarimetry. The atmospheric ozone band is masked out. The data are integrated across the central 1$\farcs$6 (21 pixels) region. 
We apply a 5-pixel boxcar smoothing to increase the signal-to-noise ratio. The resultant spectral resolution is about 48.
Overlayed red points with 1 $\sigma$-error are measurements from polarimetric imaging at 8.7,10.3, and 12.5 $\mum$.
The dashed and the dotted lines are the best fitting absorptive and emissive components, and the solid lines are the combination.  
 From upper left to the bottom right: 1) Normalized Stokes parameter \textit{q (Q/I)}. (b) Stokes \textit{u (U/I)}. (c) Polarization degree \textit{p}. 
(d) Position Angle (PA) \textit{$\theta$}. $\theta$ is approximately 20$\degr$ regardless of the wavelength. The decomposition fitting indicates that it is absorptive-polarization dominant.
}
\label{fig:4}
\end{center}
\end{figure}

For spectropolarimetry, we extracted one-dimensional spectra by integrating the central 21 pixels (1$\farcs$6) along  the  slit. We fitted a second-order polynomial of identified skylines to calibrate the wavelength. We rebinned the raw polarization data (\textit{o} and \textit{e} ray spectra) into 0.1 $\mum$  wavelength (5 pixels) bins, i.e., downsampling, and then applied an additional 5-pixel (0.1 $\mum$) boxcar-smoothing to the data to further increase the signal-to-noise ratio  resulting in the equivalent spectral resolution R$\approx$25. We masked out the region between 9.4--10.0 $\mum$, which was dominated by the atmospheric ozone features. For the total intensity spectrum presented in Fig.\,\ref{fig:2} , we smoothed the raw data with a 3-pixel (0.06 $\mum$) boxcar and the resultant spectral resolution R$\approx$50. Polarimetric spectra of WL 16 and Elias 29 are shown in Figs. \ref{fig:3} and \ref{fig:4}.

\begin{table*}
\centering
\caption{Observing Log} \label{tab:2}
\begin{tabular}{llcccccccc}

\hline
Imaging &   Filters &$\Delta\lambda$ & Date &Integration&Sensitivity&FWHM (PSF) \\

  &   ($\mu$m)  & ($\mu$m) &(UT) &Time (s)  & (mJy/10 $\sigma$/1 h) & ($\arcsec$)  \\    
\hline

            &   \textit{Si2}(8.7)   &1.1           & 2013 Aug 6  & 946       & 8.3         & 0.50   \\
            &   \textit{Si4}(10.3) &0.9           & 2013 Aug 6  &873        & 10.8       &0.60    \\ 
            &   \textit{Si6}(12.5) &0.7           & 2013 Aug 6  & 952       & 20.7    &0.40     \\
 \hline                                                 
  Spectropolarimetry   &Source(s) & &   Date      & Integration  &\\
                                   &                 & &(UT)          & Time (s)     &\\
\hline
                &WL 16      &                 &2015 Jul 4\&5  &2648& \\
                &Elias 29    &                 &2015 Jul 4\&5    &530& \\      
 \hline                               
\end{tabular} 
\end{table*}

\section{ORIGIN OF THE POLARIZATION}\label{sec:secIII}
	
Our observations (Fig.\,\ref{fig:0}) confirm that WL 16 is well-resolved and extended in the mid-IR, which was previously ascribed to the emission from PAHs and VSGs (RB03). Our spectrum (Stokes \textit{I}) of WL 16 indeed shows PAH emission features at 8.6, 11.2, 12.0, and 12.7 $\mum$ (Fig.\,\ref{fig:2}) that dominate the spectrum. Most pertinent to our primary focus on B-fields in disks is our finding that the 8.7 $\mum$ polarization vectors in WL 16 are roughly uniform in both magnitude and orientation across most of the field of view. This rough uniformity implies that the observed polarization results from grain alignment in a correspondingly uniform B-field. 
 
\subsection{Extinction toward WL 16}
\begin{figure}
\begin{center}
\includegraphics[width=\columnwidth]{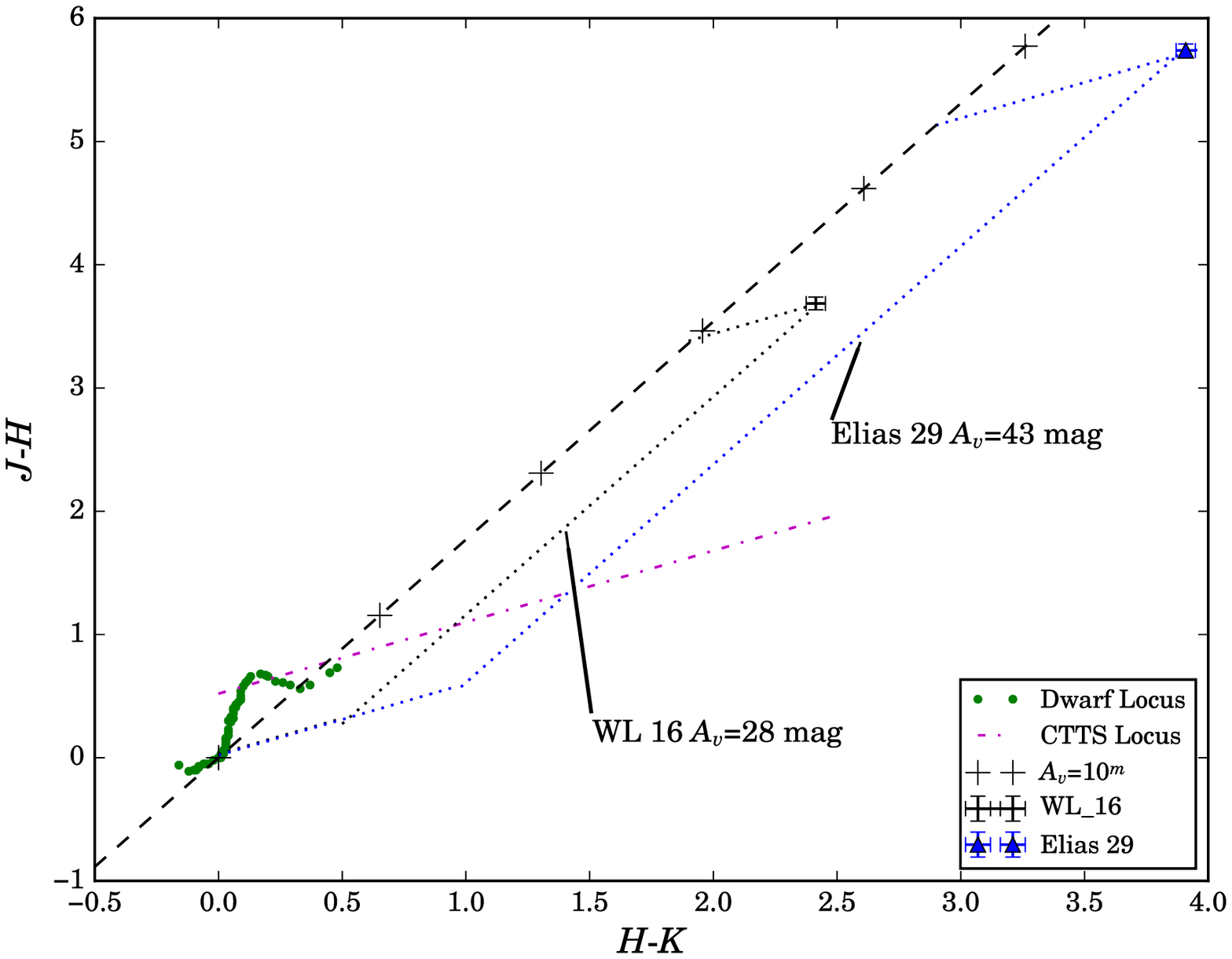}
\caption{The \textit{JHK} color-color diagram. 
The green dots are the loci of dwarfs \citep{1995ApJS..101..117K}, which represent the intrinsic color of stars of different spectral types. 
The black dashed line is the interstellar reddening vector \citep{1998AJ....115..252K}, with plus symbols indicating $A_{V}$=10 mag intervals along the vector.
The magenta dot-dashed line is the CTTS locus \citep{1997AJ....114..288M}. The black dot and blue triangle with 1 $\sigma$ errorbars are the near-IR photometry measurements of WL 16 and Elias 29 \citep{2003yCat.2246....0C}. We decompose the total extinction to the sum of an interstellar extinction vector and an intrinsic reddening vector, as shown by the dotted lines. 
 } 
\label{fig:5}
\end{center}
\end{figure}

The interstellar extinction toward WL 16 is relatively high, and our first task before characterizing the B-field giving rise to the polarization is to determine the relative roles of the polarization arising in the immediate disk environment and in the general intervening interstellar medium. The 2MASS stellar extinction map implies $A_{V}$ = 26.4$\pm$0.6 mag at the location of WL 16 \citep{2008A&A...489..143L}. 
It is not trivial to deduce the interstellar extinction for young stars, since they can also exhibit optical and near-IR excess emissions from their accreting disks. To help disentangle the interstellar extinction from the properties intrinsic to WL 16, we consider the two-color diagram as a tool to estimate the foreground interstellar extinction. \cite{1997AJ....114..288M} combined the observed near-IR properties of T Tauri stars and accretion disk models, and found that the dereddened colors of T Tauri stars occupy a narrow band, called the locus of classical T Tauri stars (CTTS locus), in the two-color (\textit{JHKL}) diagram. While recognizing that the CTTS locus is derived from T Tauri stars, and Herbig stars may have different properties, we nevertheless apply this method to WL 16 and Elias 29, taking their near-IR photometry (Table \ref{tab:4}) from \cite{2003yCat.2246....0C} and plot them on the \textit{JHK} two-color diagram in Fig.\,\ref{fig:5}. We decompose the total reddening into a component of interstellar extinction and a component of intrinsic reddening. We use the following relation to derive the corrected \textit{J}-band interstellar extinction for objects with excess in the near-IR \citep{2010ApJ...716..634G}:

\begin{eqnarray}
A_{J,{\rm corr}}&=&\frac{A_{J}/E(J-H)}
                 {k_{\rm CTTS}^{-1}-E(H-K)/E(J-H)}\nonumber\\           
               &&\left\{\frac{(J-H)-(J-H)_{0}}{k_{\rm CTTS}} 
                -\left[(H-K)-(H-K)_{0}\right]\right\}              
\end{eqnarray}

where $A_{J}$ is the total extinction, \textit{E(J-H)} and \textit{E(H-K)} are the color excesses, \textit{H-K} and \textit{J-H} are  the observed colors, \textit{(H-K)$_{0}$} and \textit{(J-H)$_{0}$} are the intrinsic colors depending on the spectral type of stars \citep{1995ApJS..101..117K}, and $k_{\rm CTTS}=0.58\pm0.11$ is the slope of CTTS locus for \textit{JHK}. Adopting the near-IR extinction relations $A_{J}/A_{K}$=2.84$\pm$0.46, \textit{E(J-H)/E(H-K)}=1.77$\pm$0.13 \citep{2014ApJ...788L..12W} and the Ophiuchus extinction law  \citep{2009ApJ...690..496C} to convert near-IR extinction to visual and mid-IR extinction, we derive a value for the interstellar extinction ($A_{V}$=28 mag) toward the central star. Thus, the interstellar material accounts for most of the extinction along our line of sight. We derive the extinction toward WL 16 at 10.3 $\mum$, $A_{10.3}$, to be 1.97$\pm$0.30 mag. The interstellar visual extinction of Elias 29 is about 43 mag and $A_{10.3}$ is 3.0$\pm$0.37 mag using the same approach (Table \ref{tab:4}).

\begin{figure}
\begin{center}
\includegraphics[width=\columnwidth]{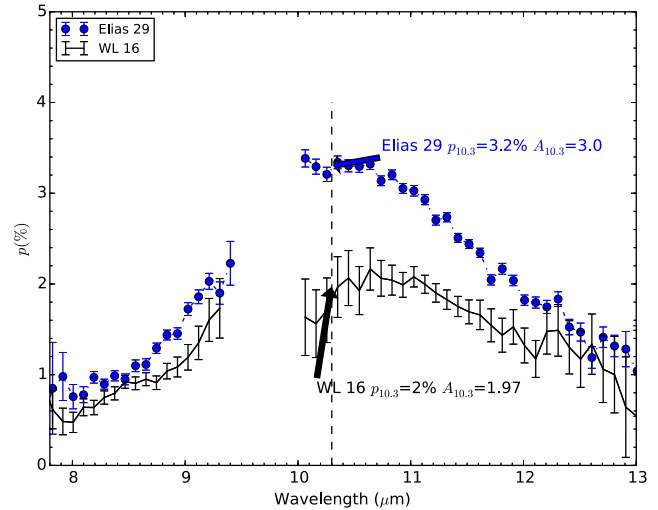}
\caption{
Comparison of the polarization profiles of WL 16 (black) and Elias 29 (blue). Both stars are located in/behind the same molecular cloud. At 10.3 $\mum$, where the polarization peaks, we find their fractional polarizations are proportional to their interstellar extinctions.}
\label{fig:6}
\end{center}
\end{figure} 

\subsection{Decomposition of Absorptive and Emissive Polarization}
We have shown that the extinction due to the foreground interstellar medium is significant, and therefore it may also account for much of the observed mid-IR polarization. However, the dust inside the disk can potentially emit polarized thermal emission that contributes to the net polarization as we see toward WL 16. To evaluation the contribution from emissive polarization, we adopted the method present in \cite{2004MNRAS.348..279A}, which showed that, in the mid-IR, the emissive and absorptive polarization components can be identified, and separated from each other, using spectral polarimetry.
Aitken's method is based on their finding that the emissive and absorptive polarization profiles ($p_{\rm em}(\lambda)$ and $p_{\rm abs}(\lambda)$) across the 10-$\mum$ silicate feature correlate with each other through the extinction, i.e., $p_{\rm em}(\lambda)=p_{\rm abs}(\lambda)/\tau(\lambda)$, where $\tau(\lambda)$ is the extinction (due to silicate) near 10 $\mum$. In practice, the absorptive profile, $p_{\rm abs}(\lambda)$, is taken as that of the Becklin-Neugebauer (BN) object, which contains absorptive polarization alone, and the extinction curve, $\tau(\lambda)$, is derived from the observations of the Trapezium region of Orion.  Aitken's method has been successfully used by a number of studies to separate emissive and absorptive polarization components from the observed total polarization when spectral polarimetry, or imaging polarimetry at multiple wavelengths, is available (e.g., \citealt{2000MNRAS.312..327S,2015MNRAS.453.2622B}).

We apply this method to the polarization data of WL 16 and Elias 29, with the results shown in Figs. \ref{fig:3} and \ref{fig:4}, respectively. Normalized Stokes parameters \textit{q} and \textit{u}, polarization fraction \textit{p}, and polarization position angle  for WL 16 and Elias 29 are shown in each plot. The position angles $\theta$ for both WL 16 and Elias 29 are almost constant with wavelength from 8.0 to 13.0 $\mum$, indicating that a single mechanism probably produces the polarization \citep{1997MNRAS.285..102E}. Should there be multiple components, we would usually expect $\theta$ to change with wavelengths (i.e., unless the absorbing and emitting fields are at 90\degr). In addition, the value of the polarization \textit{p} peaks around 10.3 $\mum$, which is expected from the absorptive polarization of astronomical-silicate feature in this spectral region. The decomposition indicates that an absorptive-dominant polarization profile provides an excellent fit for both stars, with negligible emissive components (less than 0.25\% for both cases). 
This result strengthens our conclusion that the mid-IR polarization of WL 16 is primarily due to absorption by interstellar silicate dust. Since absorption by elongated interstellar particles produces polarization parallel to the projected B-field, the direction of projected foreground magnetic field is about 33\degr$\pm$4\degr (measured from spectral polarimetry and averaged over the entire 10-$\mum$ band).

The observed polarizations at 10.3 $\mum$ for both WL 16 and Elias 29 are proportional to their respective values of interstellar extinctions as shown in Fig.\,\ref{fig:6}. From this ratio, we derive a value for the polarization efficiency, which is defined as the ratio of polarization percentage to the extinction \citep{2014ApJ...793..126C}, $(p_{10.3}/A_{10.3})\simeq1\%$ mag$^{-1}$ in Table \ref{tab:4}.  The parameter can be useful for interpreting the polarization properties of other sources and understanding the dust alignment efficiency, B-field strength, and alignment mechanisms in dense molecular clouds. This value is within the 0--3\% range derived in \cite{2000MNRAS.312..327S}. We note however that the observed polarization corresponds only to the B-field component projected on the sky (integrated along the line of sight), whereas the extinction depends on the total dust column density along the line of sight. Since there is no reason to assume that the projected B-field is the same everywhere, we recognize that the diagnostic power of this value of the ratio $(p_{10.3}/A_{10.3})$ beyond the immediate environment of WL 16 may be limited.

\begin{table}
\centering
\caption{Polarization Measurements of WL 16}
\label{tab:3}
\begin{tabular}{lcccc}
\hline
$\lambda$   &Flux Density &\textit{p}   &$\theta$    \\
  ($\mum$)          & (Jy)               &($\%$)     &($\degr$)   \\    
\hline 
8.7        &5.32(0.50)    &0.91(0.12)    &     27.09(3.71)   \\
10.3      &1.68(0.17)    &2.05(0.48)    &     29.02(6.70)    \\ 
12.5      &3.30(0.30)    &..&..                                            \\
                
\hline
\multicolumn{4}{l}{$^{ }$Values in parentheses are 1 $\sigma$ uncertainties of measurements. }\\
\multicolumn{4}{l}{$^{ }$All position angles are calibrated East from North.}

\end{tabular}
\end{table}



\begin{table}
\centering
\caption{ Extinction and Polarization of WL 16 \& Elias 29}
\label{tab:4}
\begin{tabular}{lll}
\hline
                                &WL 16   &    Elias 29 \\
\hline
\textit{J}$^{a}$         & 14.164 (0.029) & 16.788 (0.178)\\
\textit{H}$^{a}$      & 10.478 (0.023) & 11.049 (0.044)\\
\textit{K}$^{a}$      & 8.064  (0.016)  & 7.140  (0.021)\\
\textit{A$_{10.3}$}            & 1.97 (0.30) mag & 3.00 (0.37) mag \\
\textit{p$_{10.3}$}          &   2.00 (0.24)\%     &          3.21 (0.07)\%      \\
\textit{p$_{10.3}$/A$_{10.3}$} & 1.02 (0.20) \%  mag$^{-1}$ &  1.07(0.13) \% mag$^{-1}$ \\

\hline
\multicolumn{3}{l}{$^{a}$\citep{2003yCat.2246....0C}, in units of magnitude}\\
\multicolumn{3}{l}{$^{}$Values in parentheses are 1 $\sigma$ uncertainties of measurements.}
\end{tabular}
\end{table}

\subsection{Relationship of WL 16 and Regional Magnetic Fields}
\cite{1990ApJ...359..363G} probed the large-scale B-field morphology of the Ophiuchus dark cloud complex at optical wavelengths. The distribution of polarization of background starlight across this region is fairly smooth, and our polarization measurements agree well with this trend of the large-scale field (Fig. \,\ref{fig:1}). In addition, the \textit{K}-band polarimetry of WL 16 by \cite{1988MNRAS.230..321S} and \cite{2008MNRAS.384..907B}, indicates a fractional polarization \textit{p}=4.87$\pm$0.23\% with \textit{$\theta$}=27.0$\pm$1.4$\degr$ and \textit{p}=5.10$\pm$0.05\% with \textit{$\theta$}=33.90$\pm$0.03$\degr$, respectively. These values of $\theta$ are consistent with our measurements across 8--13 $\mum$, thus strengthening our conclusion that our polarization measurements do not constrain the B-field inside the disk, but instead, trace the foreground B-field.

\subsection{Intrinsic Polarization} 
While the observed polarization of WL 16 is mainly from the foreground, and the spectropolarimetry decomposition suggests a very low emissive polarization value (less than 0.25\%), can we still place a limit on the intrinsic polarization from the disk? The emissive polarization percentage detected in the case of AB Aur, an archetypal Herbig Ae star \citep{2016arXiv160902493L}, is 0.5\%. Assuming the foreground polarization is 2\% with a fixed position angle, a 0.5\% emissive polarization component could result in, at most, 7\degr deviation from the assumed foreground polarization orientation. The uncertainties of our measurements (Table \ref{tab:3} and Fig. \,\ref{fig:3}) prevent us extracting from the total observed polarization any intrinsic emissive polarization component if it is less than 0.5\%. Reiterating the limitation that we are only sensitive to the projected B-field, we place an upper limit of 0.5\% on the intrinsic polarization in WL 16. The intrinsic emissive mid-IR polarization in WL 16 is therefore likely to be much lower than a few percent predicted by \cite{2007ApJ...669.1085C}. Polarization from protoplanetary disks depends on the dust properties, e.g., dust sizes, oblateness, and composition, and also the entanglement of magnetic fields with dust grains, e.g., dust alignment and the strength of magnetic fields \citep{2013AJ....145..115H}. In the case of WL 16, it may result from a combination of these factors. 

\section{DISCUSSIONS}\label{sec:secIV}
\subsection{PAHs in WL 16}
\begin{figure}
\begin{center}
\includegraphics[width=\columnwidth]{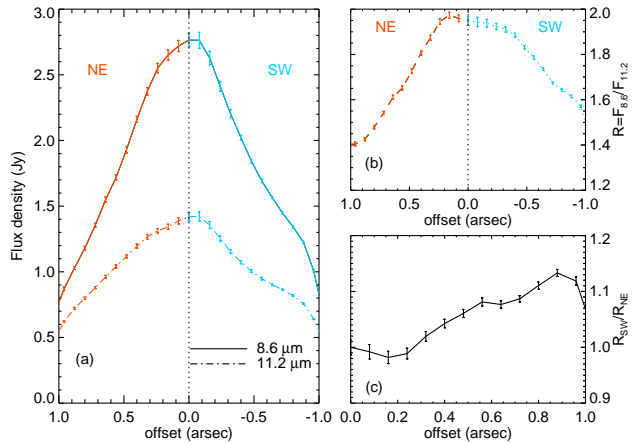}
\caption{(a): The inner 2$\farcs$0 spatial distribution of intensity at 8.6 and 11.2 $\mum$ along the major axis of the disk. There is a clear brightness asymmetry of SW (blue) and NE (red) sides of the disk at both wavelengths. (b):
\textit{R=I$_{8.6}$/I$_{11.2}$} along the major axis of the disk as a tracer of the charge state of PAHs. The larger the value, the more ionized are the PAHs. The ratio is decreasing farther from the central star as the decreasing UV radiation. (c): The relative \textit{R$_{SW}$} and \textit{R$_{NE}$}. There are more ionized PAHs at the SW side of the disk than the NE.}
\label{fig:7}
\end{center}
\end{figure}

WL 16 is rich in PAH emission features, making it useful for studying PAH properties in the environments of Herbig stars \citep{2007A&A...476..279G, 1998ApJ...504L..43D}.  ISO/SWS and Spitzer/IRS spectra of WL 16 show emission features at 6.2, 7.7, 8.6, 11.2, 12.7, 16.4, and 17.0 $\mum$ identified with de-excitation via C--C or C--H vibrational transitions of the UV-excited PAHs \citep{1984A&A...137L...5L, 2001ApJ...551..807D}. The 3.3 $\mum$ PAH feature is not detected \citep{2007A&A...476..279G}. Because a PAH molecule can be excited by a single UV photon, PAHs can trace the disk emission up to large distances from the star. The intensities of the C--C stretching and C--H in-plane bending modes, which fall in the 6--9 $\mum$ range, are generally stronger for ionized PAHs than PAH neutrals. We use the total intensity spectrum (Stokes \textit{I}) from spectropolarimetry observations to derive the ratio of the 8.6 to 11.2 $\mum$ surface brightness profiles, \textit{I$_{8.6}$} and \textit{I$_{11.2}$} (band-width 0.33 $\mum$), in order to trace the charge state of PAHs along the disk major axis \citep{1996ApJ...460L.119J}. The PAH features are often blended with neighboring PAH features (e.g., the 8.6 $\mum$  feature is on the wing of 7.7 $\mum$ feature). To correct the blended PAH emission features and subtract the continuum, we adopt the package PAHFIT by \cite{2007ApJ...656..770S} to derive the power emitted in each PAH features. There is as much as a 10\% brightness difference in the PAH features and continuum emission between the integrated NE and SW side disk flux.
Figure \ref{fig:7}a shows the derived spatial distribution of PAHs at 8.6 and 11.2 $\mum$  in  the central 2$\farcs$0 region along the major axis of the disk. Although the spatial behaviors of 8.6 and 11.2 $\mum$ are slightly different, the SW disk intensity is much lower than that of the NE disk. 
Moreover, $I_{8.6}/I_{11.2}$ reaches the maximum (about 2.0) at the central region and decreases farther from the star (Fig.\,\ref{fig:7}b), which means there are more ionized PAHs near the star, with neutral PAHs dominating the external region, as expected \citep{2001ApJ...548..296W}. 

Interestingly, we also notice (Fig.\,\ref{fig:7}b) that, while the values of \textit{I$_{8.6}$}/\textit{I$_{11.2}$} on the two  sides of the disk  are roughly comparable within 0$\farcs$3 (38 AU) of the star, they are markedly different between the two sides of the disk beyond about 0$\farcs$5 (63 AU), with \textit{I$_{8.6}$}/\textit{I$_{11.2}$} being higher on the SW side than that of the NE side in these outer regions as shown in Fig. 8c. If the disk at this large radius is optical thin in the mid-IR, the asymmetry can be explained by the neutralization of PAHs through electron recombination: PAH cations recombine more effectively with electrons in the NE than they do in the SW \citep{2003ApJ...594..987L}. 
However, WL 16, like most young Herbig-star disks (e.g., AB Aur, \cite{2006ApJ...653.1353M}), probably has an optically thick disk, the mid-IR emission that we see most likely arises in a thin layer near the disk surface (the disk `atmosphere') \citep{2014prpl.conf..339T}. Thus, the interpretation of the anti-correlation is not obvious. Complicated disk structures, such as an asymmetric puffed-up disk inner rim and disk warps, combined with the inclined viewing angle of disk, may provide explanations. To solve the puzzle, multi-wavelengths observations are necessary.      
 
We do not see any polarization features at PAH emission bands (e.g., 8.6 and 11.2 $\mum$) in the case of WL 16 (Fig. \,\ref{fig:3}). Analytical modeling does predict detectable PAH polarization in astrophysical environments \citep{2009ApJ...698.1292S}. 
Though the astrophysical conditions they considered may be different from the condition in protoplanetary disks, PAH polarization is too small (0.1--0.5\%) to be distinguished from the contribution of linear dichroism by aligned foreground dust. 

\subsection{Disk Morphology}
The total intensity maps (Fig.\,\ref{fig:0}) of WL 16 reveal intriguing asymmetric features in the disk, including: (1) an S-shaped, spiral-arm-like structure extending to both sides of the disk (green dots in Fig.\,\ref{fig:0}); (2) a dark lane immediately outside the spiral-arm-like structure in the SW disk, at 1$\farcs$3 (163 AU) from the star; (3) twisted surface brightness contours, i.e., position angles of the contours' major axes are changing from 95$\degr$ (at smaller radii) to 50$\degr$ (at larger radii); and (4) an asymmetric brightness distribution, the NE side being significantly brighter than the SW side of the disk. Though especially obvious in the PAH bands, some of these features are also seen in the continuum. In general, they appear to be qualitatively consistent with the scenario of an warped inner disk with respect to the outer disk as the result of precession induced by an unseen planet or stellar companions \citep{2015ApJ...809...93D}. In this scenario, the dark lane in the SW could be a shadow cast by the disk warp. The illumination of a warped inner disk can mimic spiral features \citep{2006ApJ...640.1078Q}. Although the twisted contours could be a result of a stellar companion of WL 16, a search for such a companion turned out to be unsuccessful \citep{2003ApJ...591.1064B}. At 40$\arcsec$ away, Elias 29 is the nearest (in projection) star of WL 16, but there is no obvious evidence that these two objects are dynamically related. We expect high-resolution and high-sensitivity radio telescopes, such as ALMA and IRAM, could lead to our new understanding of the dynamic and structure of the object.

\section{CONCLUSIONS}\label{sec:secV}
We present mid-IR polarimetric imaging and spectral observations of WL 16 obtained with GTC/CanariCam. WL 16 has a well resolved, extended disk (diameter $\simali$900 AU) in the mid-IR with $\simali$2\% polarization. Our main conclusions are as follows:
\begin{enumerate}
\item Spectropolarimetry of WL 16 firmly supports the hypothesis that the observed mid-IR polarization is dominated by absorptive polarization arising from aligned non-spherical dust grains in the foreground. Polarized emission from dust inside the disk is non-measurable with an upper limit 0.5\%. Because, in the most widely accepted dust alignment mechanisms, the absorptive polarization is parallel to the direction of B-field, we interpret the observed polarization map as indicating that the B-field in the molecular cloud is fairly uniform with projected orientation of about 33$\degr$ from North. The direction is consistent with the near-IR polarization at WL 16 and the large-scale optical $\rho$ Ophiuchus star formation region polarimetry. Though our original goal was to probe the B-field inside the protoplanetary disk, our study shows the importance of characterizing the foreground polarization as well.

\item The maximum values of the polarization of WL 16 and the nearby-polarized standard Elias 29 are proportional to their interstellar extinction. Using this ratio, we are able to characterize the polarization efficiency of dust grains in the dense molecular cloud, ($p_{10.3}/A_{10.3}$) $\simeq$1\% mag$^{-1}$. Keeping in mind that the observed polarization is associated with the projected B-field component, the parameter may be useful for constraining the dust alignment efficiency and properties in this region and for interpreting the observed polarization to other sources.

\item WL 16 is rich in PAH emission features and we have detected the 8.6, 11.2, 12.0, and 12.7 $\mum$ features in its disk. We see an asymmetry in the ratio $I_{8.6}/I_{11.2}$ between the two sides of the disk, with the NE (SW) side being brighter (fainter) at both 8.6 and 11.2 $\mum$ but with a lower (higher) value of $I_{8.6}/I_{11.2}$. This anti-correlation may be explained by complicated disk structures, e.g., warps and asymmetric disk inner rim.

\item The total intensity maps of WL 16 reveal asymmetric features, such as the S-shaped spiral-arm-like structure, twisted contours, asymmetric brightness distributions, and a dark lane on the SW side of the disk. These may indicate a disk warp, with future observations, especially with ALMA and IRAM, likely to fully clarify this picture. 

\end{enumerate}

\section*{Acknowledgements}
We thank the anonymous referee for helpful comments. We are grateful to the GTC staff for their outstanding support of the commissioning and science operations of CanariCam. C.M.T. acknowledges support from NSF awards AST-0903672, AST-0908624 and AST-1515331. E.P. acknowledges the support from the AAS through a  Chr\'{e}tien international research grant and the FP7 COFUND program-CEA through an enhanced-Eurotalent grant, and the University of Florida for its hosting through a research scholarship. C.M.W. acknowledges financial support from Australian Research Council Future Fellowship FT100100495. A.L. is supported in part by NSF AST-1109039 and NNX13AE63G.










\bsp	
\label{lastpage}
\end{document}